\documentclass{article}

\usepackage{PRIMEarxiv}

\usepackage[utf8]{inputenc} 
\usepackage[T1]{fontenc}    
\usepackage{hyperref}       
\usepackage{url}            
\usepackage{booktabs}       
\usepackage{amsfonts}       
\usepackage{nicefrac}       
\usepackage{microtype}      
\usepackage{lipsum}
\usepackage{fancyhdr}       
\usepackage{graphicx}       
\graphicspath{{media/}}     

\title{Information processing in biological molecular machines
}

\author{
  Micha\l{} Kurzy\'nski \\
  Faculty of Physics \\
  Adam Mickiewicz University \\
  Uniwersytetu Pozna\'nskiego 2 \\
  61-614 Pozna\'n, Poland\\
  \texttt{kurzphys@amu.edu.pl} \\
   \And
  Przemys\l{}aw Che\l{}miniak \\
  Faculty of Physics \\
  Adam Mickiewicz University \\
  Uniwersytetu Pozna\'nskiego 2 \\
  61-614 Pozna\'n, Poland\\
  \texttt{geronimo@amu.edu.pl} \\
}

\begin{document}
\maketitle

\begin{abstract}	
Biological molecular machines are bifunctional enzymes that catalyze two processes: one donating free energy and the other accepting it. Recent studies show that most protein enzymes have rich stochastic dynamics of transitions between the multitude of conformation substates that make up their native state. This dynamics often manifests itself in fluctuating
rates of the catalyzed processes and the presence of short-term memory. For such stochastic dynamics, after dividing the free energy into operational and organizational energy, we proved the generalized fluctuation theorem, which leads to the extension of the second law of thermodynamics to include two competing functions of process: dissipation and information. Computer simulation of the course of catalyzed processes taking place on the model network of substates, expressed in jumps of unit values at random moments of time, indicates the possibility of negative dissipation of the organizational free energy at the expense of information temporarily stored in memory, i.e. the behavior like Maxwell's demon. Because similar courses can be registered in observation of real systems, all theses of the paper are open to experimental verification. 
\end{abstract}


\section{Introduction}

The basic task of statistical physics is to combine the dynamics of the  microstates of a
studied system with the dynamics of the macrostates. The transition from microstates to
macrostates is the result of averaging over a sufficiently long period of time \cite{Penr05}.
In the statistical physics of simple systems, there are no intermediate levels of
organization between microscopic mechanics and macroscopic thermodynamics. Living matter is,
however, a complex system with the entire hierarchy of organization levels
\cite{Volk09,Eign13}. The new achievements at the turn of the century allowed us to
understand more accurately the nature of the micro and macrostates of the biological
molecular machines, which belong to the level currently being referred to as nanoscopic. 

First, the tertiary structure of the enzymatic proteins, formerly identified with one or at
most several conformational states, has been extended to the whole network of conformational
substates \cite{Karp02,Frau10,Kita98,Kurz98,Lu98,Engl06,Henz07,Uver10,Chod14,Shuk15}. The
Markov stochastic process on such a network is to be considered as the starting microscopic
dynamics for the statistical treatment. The rule are fluctuating rates of the catalyzed
reactions (the ``dynamic disorder'') \cite{Lu98,Engl06,Kurz08}, which means realization of
them in various randomly selected ways. As a symbol of the progress made recently, the
molecular dynamics study of conformational transitions in the native phosphoglycerate kinase
could be quoted, in which a network of 530 nodes was found in the long 17~$\mu$s simulation
\cite{Hu16}. This network seems to display a transition from the fractal to small-world
organization \cite{Roze10,Esco12}. On such networks, modeled by scale-free fractal trees
\cite{Goh06} extended by long-range shortcuts, the random choice of the course of catalyzed
processes is quite natural \cite{Kurz14,Kuch14}.

Secondly, new methods of stochastic thermodynamics have been applied to the description of the
non-equilibrium behavior of single nanoobjects in a finite time perspective. Work, free energy
dissipation and heat in the nanoscopic systems are random variables and their fluctuations,
proceeding forward and backward in time, proved to be related to each other by the fluctuation
theorem \cite{Evan02,Jarz97,Croo99,Seif12}. For nanoscopic machines, the concept of information
processing can be defined and the relationships between information and entropy production leads
to the generalized fluctuation theorem 
\cite{Saga10,Ponm10,Saga12,Horo13,Hart14,Horo14,Bara14,Shir15,Shir16}. It strengthened an almost
universal consensus regarding the view on the operation of  Maxwell's demon consistent with
thermodynamics \cite{Benn82,Maru09,Mand12,Luja19,Saga13,Parr15}. Thus, the demon must have a
memory and in order to use fluctuations to do the work, it reduces entropy at the expense of
gathering the necessary information on fluctuations in this memory. The information must be
erased sooner or later, and for this the same or greater work must be used.

The purpose of this paper is to answer the question of whether the particular stochastic dynamics
of the biological molecular machines allows them to act as Maxwell's demons. This would help to
understand the hard-to-interpret behavior of single-headed biological motors: myosin II
\cite{Kita99}, myosin V \cite{Wata04}, kinesin 3 \cite{Okad99}, and both flagellar \cite{Koji02}
and cytoplasmic \cite{Mall04} dynein, which can take more than one step along their track per ATP
or GTP molecule hydrolyzed. Basing on recent investigations, we assume that the stochastic
dynamics of the enzymatic machine is characterized by transient memory and the possibility to
realize the catalyzed process in various randomly selected ways. For such dynamics we calculate
both the free energy dissipation and information, and show that the first can change the sign
being partially compensated by the second. Because available experimental data on the actual
conformational transition networks in native protein machines is still very poor, we verify the
theoretical results only by computer simulations for a model network. Our conclusions are based
on analysis of the simulated time course of the catalyzed processes expressed by sequences of
jumps in the values of observed variables at random moments of time. Since similar signals can be
registered in the experiments \cite{Kita99,Wata04,Okad99,Koji02,Mall04}, all theses of the paper
are open for experimental verification. Thus, besides presentation of new theoretical concepts,
the paper can also be treated as an invitation to experimenters to perform similar analysis on
real systems.

\section{Clarification of concepts}

\subsection{Work processing}

To avoid possible confusions when using the language of stochastic thermodynamics to describe the operation of biological molecular machines, we start with a precise definition of the concepts used. For many historical reasons, the word "machine" has several different meanings in most European languages. In the context of thermodynamics, it is reasonable by a machine to understand any physical system that enables two other systems to perform work one on another. Work can be done by mechanical, electrical, chemical, thermal or other forces. Thermodynamic forces are defined as differences of the respective potentials \cite{Call85}. 

The biological molecular machines operate at a constant temperature. Under isothermal conditions,
the internal energy is uniquely divided into free energy, the component that can be converted into
work, and entropy multiplied by temperature (after Helmholtz, we call it bound energy), the
component that can be converted into heat \cite{Call85}. In view of such internal energy division, the machines are referred to as free
energy transducers \cite{Hill89}. Both thermodynamic quantities can make sense
in the non-equilibrium state if it is treated as a partial equilibrium state
\cite{Kurz06}. Besides converting work into heat or vice versa through the environment, free energy can be directly transformed into bound energy in the irreversible process of
internal entropy production which, in the energy-related context, means the free energy dissipation. 

\begin{figure}[t]
	\centering	
	\includegraphics[scale=0.8]{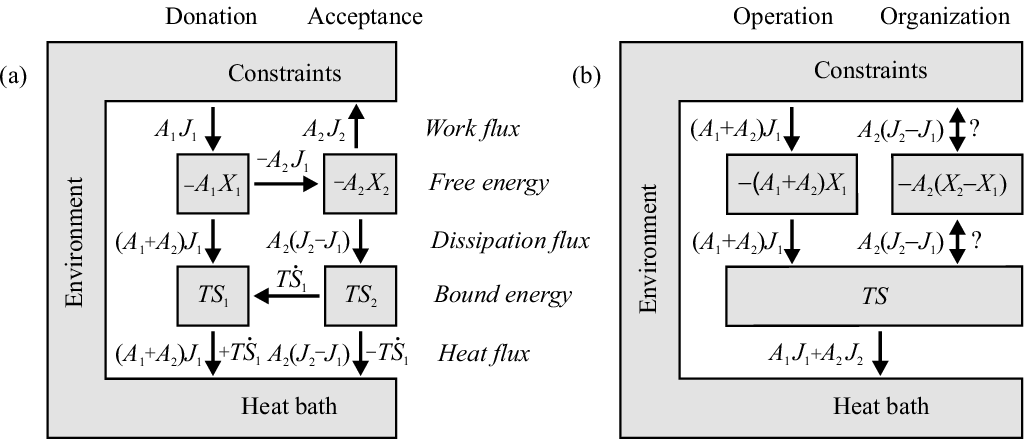} 
	\caption{Macroscopic description of the energy processing pathways in the stationary
		isothermal machine. The constraints keep stationary values of thermodynamic variables $X_1$
		and $X_2$ fixed. (a)~Division of the internal energy into free energy $F =
		-A_1X_1-A_2X_2$ (we neglect an additive constant) and bound energy $TS = TS_1+TS_2$. The directions of the energy fluxes shown are for $J_1, J_2, A_1 > 0$ and $A_2 < 0$ (in steady state, the environment compensates for incoming and outgoing fluxes). There may also be entropy transfer between subsystems. (b)~The alternative view described in the
		text. When the considered macroscopic system consists of many nanoscopic machines, their entropy is not additive and the dissipation flux $A_2(J_2-J_1)$ marked with a question mark does not have to be positive, which is discussed in this paper.}
	\label{fig1}
\end{figure}

Macroscopic description of the energy processing pathways in any
stationary operating isothermal machine is shown in Figure~\ref{fig1}a where the role of all the physical
quantities being in use is also indicated. $X_i$ denotes the input ($i=1$) or the output ($i=2$)
thermodynamic variable, $A_i$ is the conjugate thermodynamic force, and the time derivative, $J_i = \dot{X}_i$, is the corresponding flux. $S_1$ and $S_2$ are the entropies and $T$ is the temperature. To
clearly specify the degree of coupling of the fluxes, $\epsilon := J_2/J_1$, we assume that variables $X_1$ and $X_2$ are dimensionless. Corresponding forces $A_1$
and $A_2$ are then also dimensionless, if only multiplied by reversal of the thermal energy
$\beta = (k_{\rm B}T)^{-1}$ \cite{Kurz06}. We assume $J_1, J_2, A_1 > 0$ and $A_2 < 0$ throughout this paper but $-A_2 \leq A_1$, i.e., variables $X_1$ and $X_2$ are defined such that the machine does not work as a gear. In the steady state, classical Kirchhoff's law requires that the resultant work flux (the processed power) $A_1J_1 + A_2J_2$ is equal to the heat flux, and both are equal to the dissipation flux which, according to the second law of thermodynamics, must be non-negative.

From the point of view of the output force $A_2$, subsystem 1 carries out work on subsystem 2
while subsystem 2 carries out work on the environment. Jointly, the flux of the resultant work $A_2(J_2-J_1)$ is driven by the force $A_2$. The complement to
$A_1J_1+A_2J_2$ is flux $(A_1+A_2)J_1$ driven by the force $A_1+A_2$. Consequently, the free
energy processing from Figure~\ref{fig1}a can be alternatively presented as in Figure~\ref{fig1}b,
with the free energy transduction path absent. Here, two interacting subsystems of the machine
characterized by variables $X_1$ and $X_2$ have been replaced by two non-interacting subsystems
characterized by variable $X_1$ and the difference $X_2-X_1$. The first describes the ideal operation of the machine with the degree of coupling $\epsilon = 1$, while the other -- its deteriorating or improving organization. Therefore,
we will call $X_1$ the machine operational variable, and $X_2-X_1 =: X_0$ the machine organizational
variable. The first component of the total flux, $(A_1 + A_2)J_1$, achieved when the
fluxes are tightly coupled, $J_2=J_1$, must be non-negative according to the second law of thermodynamics, but
the sign of the complement $A_2(J_2-J_1)$ is open to discussion. In the macroscopic machines, the
latter is also non-negative and has the obvious interpretation of a frictional slippage in the
case of mechanical machines, a short-circuit in the case of electrical machines, or a leakage in
the case of pumps. However, in macroscopic systems composed of many nanoscopic machines, because of non-vanishing correlations
\cite{Saga10,Ponm10,Saga12,Horo13,Hart14,Horo14,Bara14,Shir15,Shir16,Mand12}, the resultant entropy $S$ is not additive as in the macroscopic machines. This could change the sign of $A_2(J_2-J_1)$ to negative, i.e., for $A_2 < 0$, to $J_2 \geq J_1$, which means that the
ratio $\epsilon = J_2/J_1$ could actually be higher than unity, as in the case of the single-headed biological motors mentioned in the Introduction \cite{Kita99,Wata04,Okad99,Koji02,Mall04}.

\subsection{Information processing}

Originally, the problem of information processing appeared in telecommunication which carried out the transmission of messages from the sender to the recipient. A message is a string of characters or, more generally, the values of some random variable. If these values are ordered in time, the message is called a signal. It is important to assume the independence of successive values of the signal, because then the signal can be treated as a statistical sample that determines the probability distribution of the random variable under consideration. The random variables ${\cal M}_1$ of the sent signal and ${\cal M}_2$ of the received signal strongly fluctuate thus their averages $M_1$ and $M_2$ do not have the same cognitive value as in thermodynamics (to distinguish random variables from their means, we mark them with uppercase calligraphic letters). More valuable are the measures of information $H_1$ and $H_2$ introduced by Shannon who, at the instigation of von Neumann, called them information entropies \cite{Cove06}. The information entropies of the sent and received signal are correlated, and the measure of this correlation is mutual information $I = H_1 + H_2 - H$, where $H$ stands for the total information entropy.

The relationship between Shannon's information entropy and the Boltzmann-Gibbs thermodynamic entropy is still hotly debated. Here, we follow the line of reasoning of  Leon Brillouin who, considering not energy $E = F + TS\,$ but entropy $S = -F/T + E/T$, introduced the term negentropy \cite{Bril56}. In order a thermodynamic system to be able to perform work under isothermal conditions, it must either collect free energy $F$ from the environment or give back free entropy $-F/T$ to it. Formally, free entropy was defined by Massieu as the Legendre transform of entropy \cite{Call85} and Brillouin understood negentropy as the negative free entropy, i.e. free energy $F$ divided by temperature $T$.

\begin{figure}[t]
	\centering
	\includegraphics[scale=0.8]{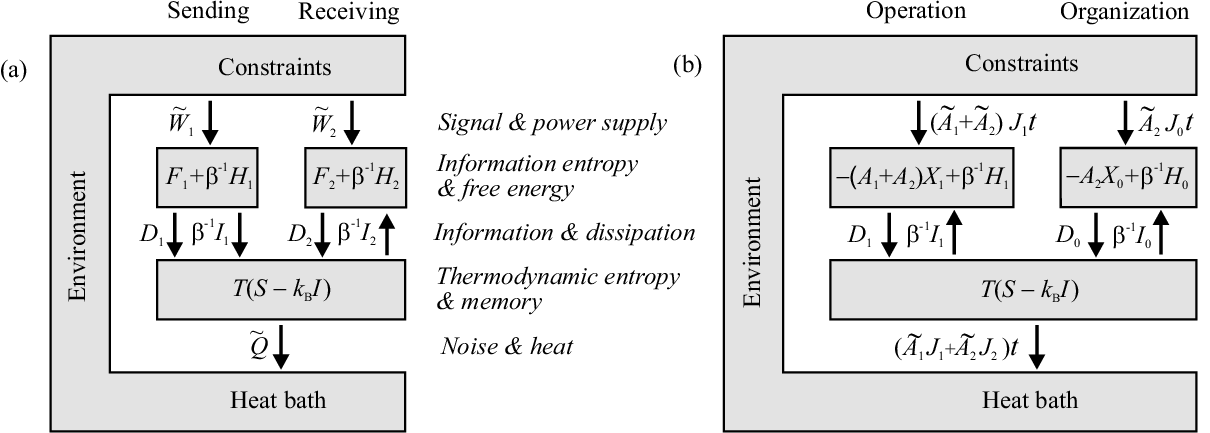} 
	\caption{Energetics of the information processing under isothermal conditions. (a) In macroscopic systems, free energies $F_1$ and $F_2$ are supplemented by Shannon's information entropies $H_1$ and $H_2$ multiplied by the molecular energy unit $\beta^{-1} = k_{\rm B}T$, while the Boltzmann-Gibbs thermodynamic entropy $S$ is enriched with memory stored in mutual information $I$. Two components $I_1$ and $I_2$ of the new function of the process, information, oppose two components $D_1$ and $D_2$ of dissipation. The macroscopic system acts as an information transmitter if $I_1 < 0$ and $I_2 > 0$. In general, the information transmitter does not operate in stationary regime, therefore the full supplied works $W_1$ and $W_2$ and the full  exchanged heat $Q$ are shown. Both work and heat fluctuate, carrying either signals or noise, which is indicated by a tilde. Since the information processor does not act as a machine in the sense of the thermodynamic definition, all the work eventually dissipates into heat but due to the small factor in the form of the molecular energy unit $\beta^{-1} = k_{\rm B}T$, dissipation is incomparably greater than information. (b)~Simultaneous processing of information and work in the isothermal nanomachine. Since information is not transduced directly, we use the work processing scheme from Figure \ref{fig1}b, where work also is not directly transduced. A detailed discussion is the subject of the following sections of this paper.}
	\label{fig2}
\end{figure}

The identification of the non-equilibrium Brillouin negentropy increment with the Shannon information entropy increment leads to the connection of thermodynamics with information theory:
\begin{equation}\label{eq30}
\Delta F/T = (W - D)/T = k_{\rm B}\Delta H \,.	
\end{equation}
Slightly different reasoning of Horowitz, Sagawa and Parrondo led to a similar relationship \cite{Horo13}.
Shannon determined the information entropy in bits using logarithms on base 2. In the context of thermodynamics, it is convenient to use natural logarithms, that is express Shannon's entropy in nats and, after multiplication by Boltzmann's constant $k_{\rm B}$, in the same units as the Boltzmann-Gibbs thermodynamic entropy.

Since the information processor does not act as a machine, the free energies of its two subsystems are independent. On the contrary, Shannon's entropies are correlated. So we have 
\begin{equation}\label{eq31}
	\Delta F = \Delta F_1 + \Delta F_2, \hspace{4pc}
	\Delta H = \Delta H_1 + \Delta H_2 - \Delta I\,.	
\end{equation}
Mutual information $I$ as well as entropies $H_1$ and $H_2$ are functions of the state of the system, while work $W$ and dissipation $D$ are functions of the process.
Both work $W$ and dissipation $D$ as increments of independent thermodynamic quantities are additive:
\begin{equation}\label{eq32}
	W = W_1 + W_2,	\hspace{4pc} D = D_1 + D_2,
\end{equation}
but in general the same cannot be said about the increase of mutual information. We have to appeal to the microscopic dynamics of the system. In the case of Markovian stochastic dynamics, only if we impose the bipartite condition, widely discussed in the applications of Shannon's entropy in the thermodynamics of nanoscopic systems \cite{Hart14,Horo14,Bara14,Shir15,Shir16}, the mutual information increment decomposes into the sum of the two independent components:
\begin{equation}\label{eq33}
	\Delta I = I_1 + I_2 .
\end{equation}
Referring to a specific realization of dynamics means that the newly introduced quantities $I_1$ and $I_2$, which we simply call informations, are functions of the process and not of the state. Moreover, since the microscopic dynamics is not known to the macroscopic observers and is hidden for them, the mutual information $I$ should be an extension of the microscopic Boltzmann-Gibbs entropy of the system. 

Considering that the non-equilibrium increment of the Boltzmann-Gibbs entropy $S$ multiplid by tempetature $T$ consists of heat $Q$ and dissipation $D$:
\begin{equation}\label{eq34}
	\Delta S = (Q + D)/T , 
\end{equation}
and substituting equations (\ref{eq31}), (\ref{eq32}) and (\ref{eq33}) to (\ref{eq30}), we find the scheme of energetics of the macroscopic information processing under isothermal conditions shown in Figure \ref{fig2}a. The important role in it is played by a new function of the process, information, with two components $I_1$ and $I_2$ opposing two components $D_1$ and $D_2$ of dissipation. 

In macroscopic telecommunication (Figure \ref{fig2}a), the signals carrying information require amplification and conversion (recording or modulation of a specially generated carrier wave as well as possible encoding/decoding) at the expense of the work provided by the appropriate power supplies. All this work eventually dissipates into heat. Just as work fluctuations are the source of signals delivered to or received from the system, heat fluctuations are the source of noise. Due to the small factor in the form of the molecular energy unit $\beta^{-1} = k_{\rm B}T$, dissipation is incomparably greater than information. Only in nanoscopic systems, in particular in nanomachines (Figure \ref{fig2}b), dissipation can compete with information, which will be discussed in detail in Section 4.

\section{Biological molecular machines as chemo-chemical nanomachines}

\begin{figure}[h]
	\centering
	\includegraphics[scale=0.65]{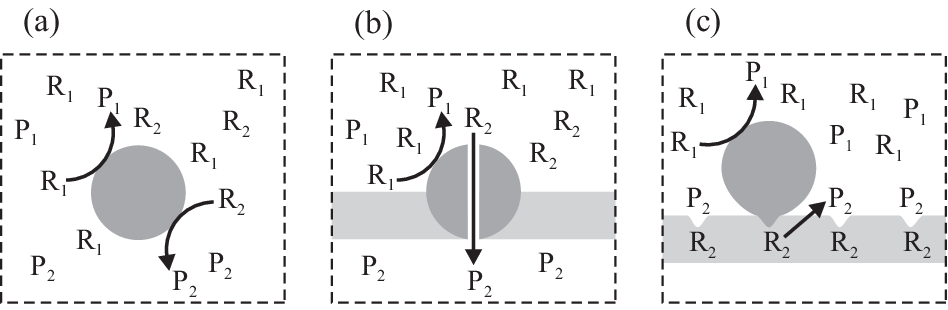}
	\caption{Schematic picture of the three types of the biological molecular machines.
		(a)~Enzymes that simultaneously catalyze two reactions. (b)~Pumps placed in a biological
		membrane -- the molecules present on either side of the membrane can be considered to occupy
		different chemical states. (c)~Motors moving along a track -- the external load influences the	free energy of motor binding, which can be considered as a change in the probability of
		finding a particular track's binding site, hence its effective concentration \cite{Kurz06,Kolo07}. Dashed lines represent constraints, the operation of which is described in the text. It should be emphasized that the entire system in the frame is a machine: these are the constraints that determine the values of the thermodynamic variables and forces.}
	\label{fig3}
\end{figure}

Since the biological molecular machines operate by thermal fluctuations, just like chemical reactions, we
treat them as chemo-chemical machines \cite{Kurz06}. The protein chemo-chemical machines can be
considered as bifunctional enzymes that simultaneously catalyze two effectively unimolecular
chemical reactions: the free energy-donating input reaction ${\rm R}_1 \rightarrow {\rm P}_1$ and the free
energy-accepting output reaction ${\rm R}_2 \rightarrow {\rm P}_2$, see Figure~\ref{fig3}a. Also pumps and molecular motors can be treated in the same way, see Figures~\ref{fig3}b and \ref{fig3}c. Macroscopically, with some relationships between the concentrations of all reactants
\cite{Kurz06}, non-equilibrium molar concentrations of product molecules [P$_1$] and [P$_2$]
related to enzyme total concentration [E] are the two dimensionless thermodynamic variables $X_1$
and $X_2$ determining the steady state of the whole system of biological molecular machines.
These, along with the conjugate thermodynamic forces, i.e. chemical affinities $A_1$ and $A_2$
for reactions ${\rm R}_1 \rightarrow {\rm P}_1$ and ${\rm R}_2 \rightarrow {\rm P}_2$,
respectively, determine work performed on and by the machines. Assuming the molecule solution
is perfect, the formal definitions take the form \cite{Hill89,Kurz06}:
\begin{equation}
	\label{eq1}
	X_{i}:=\frac{[\mathrm{P}_{i}]}{[\mathrm{E}]}\,,\hspace{4pc}
	\beta A_{i}:=\ln\frac{[\mathrm{P}_{i}]^{\mathrm{eq}}}{[\mathrm{R}_{i}]
		^{\mathrm{eq}}}\frac{[\mathrm{R}_{i}]}{[\mathrm{P}_{i}]}\,,
\end{equation}
$i = 1, 2$. $\beta$ is the reciprocal of the thermal energy $k_{\rm B}T$ and the superscript eq denotes the
equilibrium concentrations.

The thermodynamics of a single nanoscopic system is based on the fluctuation theorem which describes the response of the system to a "protocol" determined
by adopting certain control parameters \cite{Evan02,Jarz97,Croo99,Seif12}. The nanoscopic machine
we consider consists of a single enzyme macromolecule surrounded by a solution of its substrates
and products. The whole is an open non-equilibrium system which operates stochastically under the
influence of external control parameters whose role in the steady state is fulfilled by the constant thermodynamic forces $A_1$ and $A_2$. For a single nanomachine, fixing the forces $A_1$ and $A_2$ and the variables $X_1$ and $X_2$ defining them following Equations (\ref{eq1}) is formally equivalent to keeping the machine in the steady state by imaginary constraints (see Figure ~\ref{fig3}) that control the number of incoming and outgoing product molecules of both catalyzed reactions to eliminate the effect of their formation and annihilation in the machine turnovers. It looks as if two signals pass through the constraints, in the form of two sequences of bits or rather signs, e.g. $\dots, +, +, -, +, -, +, +, +, -, -, +, \dots\,$ for the subsequent formation or annihilation of the product molecules. For a thermodynamic nanoscopic observer, the statistically independent random variables of the signals are the resultant numbers of molecules P$_1$ and P$_2$ created in successive equal, sufficiently long but finite periods of time $t$. They are determined by the fluxes ${\cal J}_1(t)$ and ${\cal J}_2(t)$, respectively, conjugate to the forces $A_1$ and $A_2$, limited to the time interval $t$ and multiplied by this time. The products ${\cal J}_1(t)t$ and ${\cal J}_2(t)t$ correspond to the random variables ${\cal M}_1$ and ${\cal M}_2$ considered in the Section 2.2 in the definition of Shannon's information entropy.

\begin{figure}[t]	
	\centering
	\includegraphics[scale=0.6]{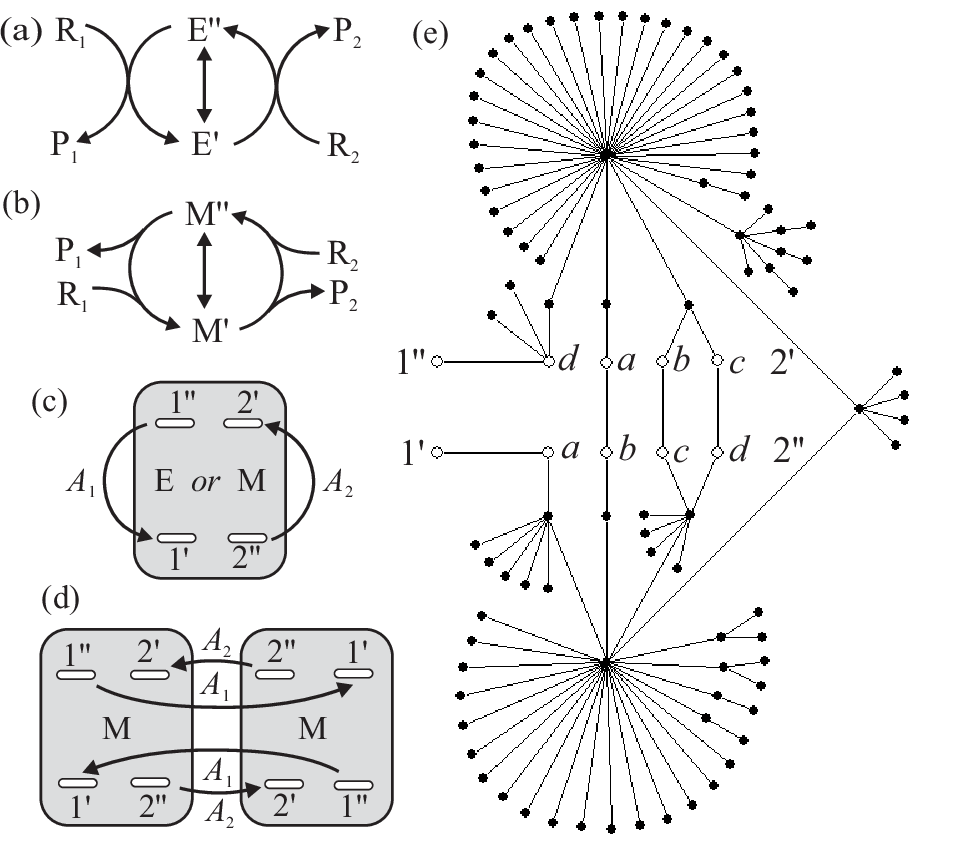}
	\caption{Dynamics of the enzymatic chemo-chemical machine. (a)~and (b)~Schemes of
	conventional chemical kinetics that take into account only two states E$''$ and E$'$ of the
	free enzyme or the two states M$''$ and M$'$ of the enzyme-reagents complex. (c)~The
	generalized scheme \cite{Kurz14,Kurz03}. The gray box represents a network of
	transitions between conformational substates of the enzyme or the enzyme-reactants complex
	native state. All these transitions satisfy the detailed balance condition. A system of single
	or multiple (ovals) pairs of conformational substates (the "gates") $(1'', 1')$ and $(2'',
	2')$ has been distinguished, between which input reaction ${\rm R}_1 \rightarrow {\rm P}_1$
	and output reaction ${\rm R}_2 \rightarrow {\rm P}_2$ cause, respectively, additional
	transitions that break the detailed balance. All the reactions are reversible; the arrows
	indicate the directions assumed to be forward. (d)~Bipartite structure of the formally doubled
	network described in the text. (e)~Sample realization of the 100-node network, constructed
	following the algorithm described in Section 5.}
	\label{fig4}
\end{figure}

In the language of conventional chemical kinetics, the operation of a chemo-chemical machine is
explained by the scheme in Figure~\ref{fig4}a: the energy-donating reaction ${\rm R}_1 \rightarrow
{\rm P}_1$ forces the direction of the energy-accepting reaction ${\rm R}_2 \rightarrow {\rm
P}_2$, though the nonequilibrium concentration values [R$_2$] and [P$_2$] would prefer the
opposite direction \cite{Hill89,Kurz06}. We assumed that the binding of reactants precedes their
detaching but this order may be reversed as shown in Figure~\ref{fig4}b. In both cases, for the
sake of simplicity, we do not take into account details of the binding and detaching reactions
involving spatial diffusion of reactants and molecular recognition \cite{Kurz06,Kurz03}. 

Only two states E$'$ and E$''$ of the free enzyme are distinguished in the scheme in
Figure~\ref{fig4}a and only two states M$'$ and M$''$ of the enzyme-reagents complex are
distinguished in the scheme in Figure~\ref{fig4}b. Only one non-reactive transition between them
is assumed. However, the nanoscopic kinetics of biological
machines is described by the whole network of conformational substates. In this network, with
transitions that obey the detailed balance condition, a system of pairs of nodes (the ``gates'')
is distinguished, between which the input and output chemical reactions force additional
transitions that break the detailed balance in a degree dependent on the values of $A_1$ and $A_2$
\cite{Kurz14,Kurz06,Kurz03}, see Figure~\ref{fig4}c. The rule is the presence of many gates
allowing the choice of the way of implementing reactions \cite{Lu98,Engl06}.

The scheme in Figure~\ref{fig4}c determines the $J_1$ and $J_2$ fluxes in the statistical ensemble of many molecular machines, but does not envisage the operation of an imaginary microscopic observer placed on the constraints of a single machine to pass through them the molecules P$_1$ or P$_2$ sequentially created or destroyed by the machine. Including the observer's memory in this scheme would mean non-Markovianity
of the process, but this can be avoided by formal doubling the network
\cite{Bara14} and replacing the scheme in Figure~\ref{fig4}c with that
presented in Figure~\ref{fig4}d. Now, the state of the machine is determined not only by the
network node $s$, but also by the end node $g$ of the last gate passed. All the transitions
within the doubled network in Figure~\ref{fig4}d can be divided into the internal transitions of the
form $(g,s) \to (g,s')$ (the index $g$ of the end node of the last gate passed remains the same)
and the external transitions of the form $(g,s) \to (g',g')$ (the last gate passed turns into a new one, the end node of which simultaneously becomes the initial node for further internal passages). Since external transitions are realized by either a type 1 input gate or a type 2 output gate, the external transition network has the bipartite structure
\cite{Hart14,Horo14,Bara14,Shir15,Shir16}.

All the internal transitions are hidden for the nanoscopic observer, who observes not even individual external transitions, but only the resultant number of particles P$_1$ and P$_2$ created in some equal consecutive time intervals $t$ with their combined effect in the form of fluxes ${\cal J}_1(t)$ and ${\cal J}_2(t)$ resulting from averaging the stochastic trajectory in the state space from Figure~\ref{fig4}d.

\section{Information versus dissipation}

The consequence of distinguishing the external and internal stochastic dynamics of a nanoscopic machine is the division of the network on which this dynamics is realized into two subnetworks: the observable and the hidden one \cite{Horo14,Bara14,Shir15,Shir16}. Separately, the dynamics on the hidden subnetwork satisfies the detailed balance condition, while the dynamics on the observable subnetwork is controlled by external forces. In addition to free energy processing, the result of the joint action of both dynamics is free energy dissipation, specified by the Jarzynski equality \cite{Shir15,Shir16}: 
\begin{equation}
	\label{eq2}
	\langle\exp(-\sigma)\rangle = 1.
\end{equation}

The nanomachines operate in a stationary regime macroscopically characterized by Figure~\ref{fig1}. However, in the case of a single nanomachine, the observed fluxes, determined in selected finite time intervals $t$, are random variables. It follows that 
the stochastic dimensionless entropy production $\sigma$, equal to free energy dissipation divided by the thermal energy  $k_{\rm B}T = \beta^{-1}$, takes the form  
\begin{equation}
	\label{eq3}
	\sigma = \beta[A_1\mathcal{J}_1(t) + A_2\mathcal{J}_2(t)]t =
	\beta[(A_1 + A_2)\mathcal{J}_1(t) + A_2\mathcal{J}_0(t)]t \,.
\end{equation}
We wrote two alternative divisions of the dissipation flux discussed in Figure~\ref{fig1} and all further considerations may equally apply to either of the divisions but, because
the two information subsystems in Figure~\ref{fig2}a do not interact directly with each other, we will continue the discussion for directly non-interacting fluxes ${\cal J}_1(t)$ and ${\cal J}_0(t) = {\cal J}_2(t) - {\cal J}_1(t)$. 

For the second division in Equation (\ref{eq3}), the detailed fluctuation theorem corresponding to the integral fluctuation theorem in Equation~(\ref{eq2}) is
\begin{equation}
	\label{eq5}
	\frac{p(j_{1},j_{0})}{p(-j_{1},-j_{0})}
	= \exp \sigma(j_1,j_0) = \exp \beta[(A_{1}+A_{2})j_{1}+A_{2}j_{0}]t \,.
\end{equation}
$j_1$ and $j_0$ in Equation~(\ref{eq5}) denote particular values of random fluxes $\mathcal{J}_1(t)$ and $\mathcal{J}_0(t)$, respectively, $p$ is the joint probability distribution function
of these fluxes and the average in Equation~(\ref{eq2}) is performed over this distribution. By
definition, the stationary averages of the random fluxes are time independent,
$\langle\mathcal{J}_i(t)\rangle = J_i$
for any properly selected value of $t$, but the values of the higher moments of
distribution $p(j_{1},j_{0})$ increase with the shortening of time $t$. The convexity of
the exponential function provides the second law of thermodynamics
\begin{equation}
	\label{eq6}
	\langle\sigma\rangle = \beta [(A_1+A_2)J_1 + A_2J_0] t \geq0
\end{equation}
to be a consequence of the Jarzynski equality in Equation~(\ref{eq2}).

Equation~(\ref{eq5}) is identical to the fluctuation theorem for the stationary fluxes earlier derived by Schnakenberg, Andrieux and Gaspard \cite{Schn76,Andr07,Gasp13} who based it on much more complex
considerations finished with the limit $t \to \infty$. Here we can consider different options for choosing time $t$, which must be long enough for the considered statistical ensemble to comprise only
stationary fluxes but finite to observe any fluctuations. As all stationary fluxes in the ensemble
are statistically independent, the probability distribution function $p(j_1,j_0)$ is the
two-dimensional Gaussian~\cite{Sirc16}
\begin{eqnarray}
	\label{gauss}
	p(j_1,j_0) &=& \frac{1}{2\pi\mathit{\Delta}_1\mathit{\Delta}_0\sqrt{1-\rho^2}}
	\exp \left\{-\frac{1}{2(1-\rho^2)}\right. 
	\left[\frac{(j_1-J_1)^2}{\mathit{\Delta}_1^2}\right. \nonumber\\
	&-&2 \rho\frac{(j_1-J_1)(j_0-J_0)}{\mathit{\Delta}_1\mathit{\Delta}_0}
	\left.\left.+\frac{(j_0-J_0)^2}{\mathit{\Delta}_0^2}\right]\right\} \,,
\end{eqnarray}
where the averages $J_i = \langle\mathcal{J}_i(t)\rangle$,  $i = 1, 0$, and the  corresponding
variances (the squares of the standard deviations) $\mathit{\Delta}_i^2 := \langle(\mathcal{J}_i(t) -
J_i)^2\rangle$ specify the Gaussian marginals for the individual fluxes, and $\rho$ is the
correlation coefficient:
\begin{equation}
	\label{eq8}
	\rho:=\frac{\left<(\mathcal{J}_1(t)-J_1)(\mathcal{J}_0(t)-J_0)\right>}
	{\mathit{\Delta}_1\mathit{\Delta}_0} \,.
\end{equation}
Here and further on, for the sake of brevity, we omit argument $t$ specifying $\mathit{\Delta}_1$, $\mathit{\Delta}_0$ and $\rho$. The requirement for the Gaussian in Equation~(\ref{gauss}) to satisfy the detailed fluctuation theorem
in Equation~(\ref{eq5}) leads to a system of two quadratic equations that link standard deviations
$\mathit{\Delta}_1$ and  $\mathit{\Delta}_0$ with averages  $J_1$ and $J_0$:
\begin{equation}
	\label{eq9}
	\mathit{\Delta}_1^2 + \frac{A_2}{A_1+A_2}\rho
	\mathit{\Delta}_0\mathit{\Delta}_1=\frac{2J_1}{\beta(A_1+A_2)t} \,, \hspace{4pc}
	\mathit{\Delta}_0^2 + \frac{A_1+A_2}{A_2}\rho
	\mathit{\Delta}_0\mathit{\Delta}_1
	=\frac{2J_0}{\beta A_2t} \,.
\end{equation}
From Equations~(\ref{eq9}), it follows that standard deviations $\mathit{\Delta}_1$ and 
$\mathit{\Delta}_0$ are inversely proportional to the square root of $t$. For the lack of
correlations, $\rho = 0$, the solutions to Equations~(\ref{eq9}) reconstruct our earlier result
\cite{Chel17}.

Equations~(\ref{eq9}) allow to calculate the ratios of marginals $p(j_1)/p(-j_1)$ and 
$p(j_0)/p(-j_0)$  for fluxes $\mathcal{J}_1(t)$ and $\mathcal{J}_0(t)$ separately, from which it
follows that, after logarithming, the detailed fluctuation theorem for the flux $\mathcal{J}_1(t)$ takes the form
\begin{equation}
	\label{eq10}
	\ln\frac{p(j_{1})}{p(-j_{1})} = \beta(A_{1}+A_{2})j_{1}t 
	+ \rho \frac{\mathit{\Delta}_0}{\mathit{\Delta}_1} \beta A_{2} j_{1}t
	=: \sigma_1(j_1) - \iota_1(j_1)\,, 
\end{equation}
and for the flux $\mathcal{J}_0(t)$ -- the form
\begin{equation}
	\label{eq11}
	\ln\frac{p(j_{0})}{p(-j_{0})} = \beta A_{2} j_{0}t 
	+ \rho \frac{\mathit{\Delta}_1}{\mathit{\Delta}_0} \beta(A_{1}+A_{2}) j_{0}t 
	=: \sigma_0(j_0) - \iota_0(j_0) \,.
\end{equation}
The corresponding integral fluctuation theorems are
\begin{equation}
	\label{eq12}
	\langle\exp(-\sigma_1+\iota_1)\rangle = 1 \,, \hspace{4pc}
	\langle\exp(-\sigma_0+\iota_0)\rangle = 1 \,.
\end{equation}
The averages are performed over the one-dimensional distributions, either $p(j_1)$ or $p(j_0)$. 

In addition to the functions $\sigma_1$ and $\sigma_0$ that describe separate entropy productions,
two new functions $\iota_1$ and $\iota_0$ have appeared. As $\mathit{\Delta}_1$ and
$\mathit{\Delta}_0$ are positive and forces $\beta (A_1+A_2)$ and $\beta A_2$ are of the opposite
signs, components $\sigma_i$ and $\iota_i$ subtract when $\rho$ is positive and add up when $\rho$
is negative. To interpret both the new functions, let us add their mean values $\langle \iota_1 \rangle $ and $\langle \iota_0 \rangle$ together. From Equations~(\ref{eq5}), (\ref{eq10}) and (\ref{eq11}), the
relationship
\begin{equation}
	\label{eq13}
	\langle \iota_1 \rangle + \langle \iota_0 \rangle = \left\langle
	\ln\frac{p(\mathcal{J}_1,\mathcal{J}_0)}{p(\mathcal{J}_1)p(\mathcal{J}_0)} - \ln\frac{p(-\mathcal{J}_1,-\mathcal{J}_0)}{p(-\mathcal{J}_1)p(-\mathcal{J}_0)} \right\rangle 
\end{equation}
results. The average of the first logarithm represents the mutual information between the fluxes $\mathcal{J}_1(t)$ and $\mathcal{J}_0(t)$ that go forward, whereas the average of the second logarithm -- between those fluxes that go backward. The difference determines the mutual information increment caused by the transfer of information from the operational to the organizational subsystem. Referring to bipartite structure of the external transition network, for the biological nanomachines stated in Section 3, and comparing Equation~(\ref{eq13}) with Equation~(\ref{eq33}), we find that $\langle \iota_1 \rangle$ and $\langle \iota_0 \rangle$ should be interpreted as the informations sent, respectively, from the operational and the organizational subsystems of the free energy to the bound energy. In consequence, bound energy is what constitutes the information reservoir \cite{Bara14,Shir15,Shir16,Mand12,Luja19,Deff13}. Let us emphasize that because $\iota_1$ is the function of only $j_1$, and $\iota_0$ is the function of only $j_0$, the value of the difference of the logarithms in Equation~(\ref{eq13})
remains the same when averaged over the correlated distribution $p(j_1,j_0)$ or the product
$p(j_1)p(j_0)$. Of course, this is not true for the separate components of the logarithms difference in Equation~(\ref{eq13}).
 
With the help of Equations~(\ref{eq9}), Equations~(\ref{eq10}) and (\ref{eq11}) can be rewritten directly as
\begin{equation}
	\label{eq14}
	\ln\frac{p(j_{1})}{p(-j_{1})} = \frac{2J_1}{\mathit{\Delta}_1^2} j_1
	=: \beta (\tilde{A_1}+\tilde{A_2}) j_1 t
\end{equation}
for the flux $\mathcal{J}_1(t)$ and 
\begin{equation}
	\label{eq15}
	\ln\frac{p(j_{0})}{p(-j_{0})} = \frac{2J_0}{\mathit{\Delta}_0^2} j_0
	=: \beta \tilde{A_2} j_0 t
\end{equation}
for the flux $\mathcal{J}_0(t)$. Compared to forces $A_1$ and $A_2$, the forces $\tilde{A}_1$ and $\tilde{A}_2$ defined above contain corrections for signals penetrating the constraints in the form of fluctuating fluxes ${\mathcal J}_1(t)$ and ${\mathcal J}_2(t)$. As in the case of information processing in macroscopic systems shown in Figure~\ref{fig2}a, this corresponds to an increase of free energies by Shannon's information entropies. In general, the corrections to forces $A_1$ and $A_2$ play the role of additional chemical forces acting on each of the stationary fluxes separately \cite{Horo13}. On introducing the averaged values
\begin{equation}
	\label{eq16}
	\langle {\cal J}_i(t)\rangle = J_i \,, \hspace{4pc}
	\langle\sigma_i\rangle =: \beta D_i \,, \hspace{4pc} 
	\langle\iota_i\rangle =: I_i \,, \hspace{4pc}  i = 1, 0\, ,
\end{equation}
we obtain the image of simultaneous processing of information and work in the isothermal nanomachine, shown at the beginning in Figure~\ref{fig2}b.

From the integral fluctuation theorems in Equations~(\ref{eq12}), it follows that dissipations and informations satisfy the
generalized second laws of thermodynamics:
\begin{equation}
	\label{eq17}
	\beta D_1 \geq I_1 \,, \hspace{4pc} \beta D_0 \geq I_0 \,.
\end{equation}
$D_1$ is always positive, but if $I_0$ is negative, also $D_0$ can be negative, which means that the organizational subsystem may behave like Maxwell's demon. Unfortunately, the general theory does not provide for signs $D_0$, $I_1$ and $I_0$. Because contributions from dissipation $D_1$ and $D_0$ depend only on the values of stationary fluxes $J_1$ and $J_2$, they can be determined macroscopically for specific values of forces $A_1$ and $A_2$ \cite{Kurz14}. More detailed statistics are needed to determine contributions from informations $I_1$ and $I_0$. A comparison of two equivalent formulations of the fluctuation theorem for separate fluxes,
Equations~(\ref{eq10}) and (\ref{eq11}) with Equations~(\ref{eq14}) and (\ref{eq15}), leads to
expressions for stochastic informations $\iota_1$ and $\iota_2$ dependent only on standard
deviations $\mathit{\Delta}_1$ and $\mathit{\Delta}_0$ but not on the correlation coefficient
$\rho$. Only four parameters  $J_i$ and  $\mathit{\Delta}_i$,  $i = 1, 0$, that characterize the
Gaussian distribution functions of the separate fluxes, are sufficient to specify the averages $I_1$ and $I_0$. But for that we need to know the microscopic dynamics of the nanomachine.

\section{Work and information processing in a model system}

\subsection{Specification of the model}

Because available experimental data on the actual conformational transition networks in native
proteins are still very poor, we constructed a model network to apply the theory for interpreting
results of computer simulations. As mentioned in the Introduction, this network should be
scale-free and display a transition from the fractal to small-world organization
\cite{Hu16,Roze10,Esco12,Goh06,Kurz14}. A sample network of 100 nodes with such property is
depicted in Figure~\ref{fig4}e. The algorithm of constructing the stochastic scale-free fractal
trees was taken from Goh et al. \cite{Goh06}. Shortcuts, though more sparsely distributed, were
considered after Rozenfeld, Song and Makse \cite{Roze10}. Here, we randomly chose three shortcuts
from the set of all the pairs of nodes distanced by six links. The network of 100 nodes in
Figure~\ref{fig4}e is too small to determine its scaling properties, but a similar procedure of
construction applied to $10^5$ nodes results in a scale-free network, which is fractal on a small
length-scale and a small world on a large length-scale. 

To equip the network with stochastic dynamics, we assume the probability of changing a node to
any of its neighbors to be the same in each random walk step \cite{Kurz14,Chel17}. Then,
following the detailed balance condition, the equilibrium occupation probability of a given node is inversely
proportional to the number of its links (the node degree). The most stable nodes are the hubs. The
assumed dynamics is independent of temperature, and therefore not very realistic, but this problem
is not the subject of our research here. For given node $l$, the transition probability to any of
the neighboring nodes per random walk step is one over the number of links and equals
$(p_l^{\rm eq}\tau_{\rm int})^{-1}$, where $p_l^{\rm eq}$ is the equilibrium occupation
probability of the given node and $\tau_{\rm int}$ is the mean time to repeat a chosen internal
transition, counted in the random walk steps. This time is determined by the doubled number of
links minus one \cite{Chel17}, $\tau_{\rm int}=2\cdot(100+3-1)=204$ random walk steps for the
100 node tree network with 3 shortcuts assumed.

In the scheme in Figure~\ref{fig4}e, let us note two hubs, the most stable protein substates, usually the only ones occupied sufficiently high to be observable under equilibrium conditions and identified as, e.g., ``open'' and ``closed'', or ``bent'' and ``straight''. The proposed kind of
stochastic dynamics combines in fact a chemical description with a mechanical description that
takes into account randomness of the mechanical movements \cite{Kurz06}. 

For simplicity, a single input gate is assumed. A single output gate chosen for the simulations
is created by the pair of transition states $(2''a, 2'd)$. The four pairs of transition states
$(2''a, 2'a)$, $(2''b, 2'b)$, $(2''c, 2'c)$ and $(2''d, 2'd)$ lying tendentiously one after
another, create alternative fourfold output gate. The external transition probability competes
with the probability of possible internal transitions and, for the forward transition, equals
$(p_i^{\rm eq}\tau_{\rm ext})^{-1}$ per random walk step, $p_i^{\rm eq}$ denoting the equilibrium
occupation probability of the initial input or output node ($i=1,2$, respectively). The
corresponding backward external transition probability is modified by detailed balance breaking
factors $\exp (-\beta A_i)$. For the mean time of forward external transitions we assume $\tau_{\rm ext}=20$ random walk steps, one
order of the magnitude shorter than for the mean time of internal transitions $\tau_{\rm int}$, which
means that the whole process is controlled by the internal dynamics of the system. The identification of the time unit with the computer step (CPU clock rate) still requires the determination of the probabilities of
stay, different in network and gate nodes \cite{Chel17}.

\subsection{Results of computer simmulations}

\begin{figure}[h!]
	\centering
	\includegraphics[scale=0.5]{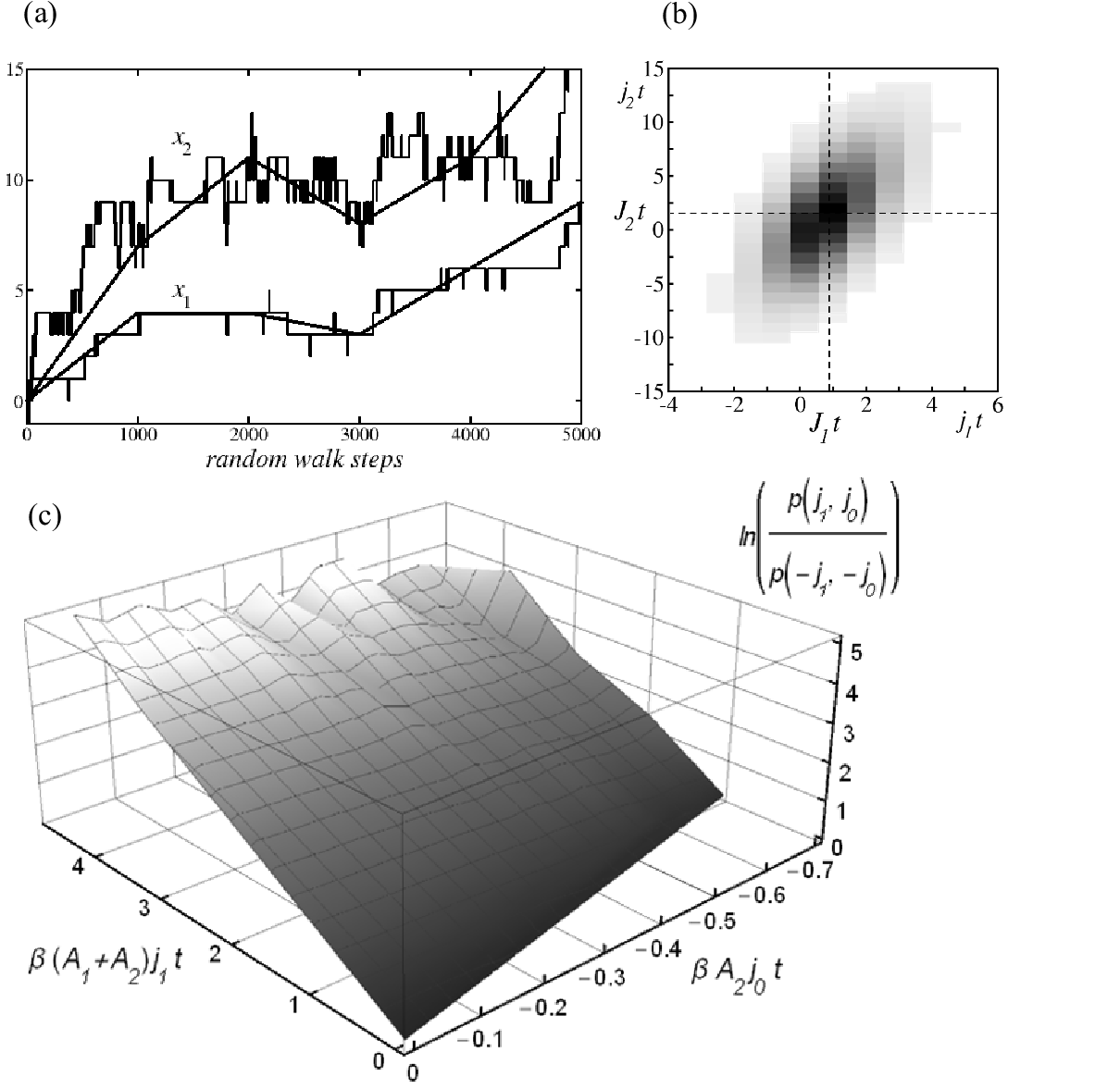}
	\caption{(a)~Simulated time course of the resultant number of external transitions $x_1$ and $ x_2$
		through the input and output gates, respectively, produced in the successive random walk steps on the
		network presented in Figure~\ref{fig4}e with $\beta A_1 = 1$, $\beta A_2 = -0.05$ and the
		fourfold output gate. Determination of the corresponding fluctuating fluxes $j_1$ and $j_2$ by
		averaging over successive time periods of $t = 1000$ steps is shown. (b)~The  determined
		Gaussian distribution of the fluxes $j_1$ and $j_2$. (c)~Numerical verification of the
		detailed fluctuation theorem in Equation~(\ref{eq5}). Deviations from the plane for large
		values of $j_1t$ and $j_0t$ result from insufficient statistical material collected in
		simulations.}
	\label{fig5}
\end{figure}

To check if the organizational subsystem of the biological molecular machine can actually behave like Maxwell's
demon, we performed computer simulations of random walk on the network constructed following the
algorithm described above. We assumed $\beta A_1 = 1$ and a few smaller, negative values of $\beta A_2$. 
The result of each
simulation were the time courses of the resultant numbers $x_1$ and $x_2$ of passes through
input and output gates, respectively. Exemplary courses are shown in Figure~\ref{fig5}a. Assuming renewal character of time intervals between jumps, the time courses $x_1$ and $x_2$ are continuous time random walks
\cite{Sirc16,Metz14}. The thermodynamic description of the nanosystem requires averaging over a finite observation
period. To get statistical ensembles of the stationary fluxes, we divided long stochastic
trajectories of $10^{10}$ random walk steps into segments of equal lengths $t$. For the fluxes $j_i$
to be statistically independent, the selected time $t$ must have been longer than the complete cycle of free energy transduction. By analyzing the dwell time distribution of cycles, we found the reasonable time of
averaging to be $t = 1000$ random walk steps. The result of averaging are pairs of fluxes $j_i =
\Delta x_i/t$, $i = 1, 2$, which are also shown in Figure~\ref{fig5}a.

The distribution of random fluxes $j_1$ and $j_2$ found for $\beta A_1 = 1$, $\beta A_2 =
-0.05$ and the fourfold output gate is depicted in Figure~\ref{fig5}b. The distribution is
actually Gaussian, hence the thermodynamic behavior of a single biological nanomachine for a fixed averaging period $t$ is free diffusion around a point with coordinates $(J_1t, J_2t)$. After exchanging the variables from $j_1$ and $j_2$
to $j_1$ and $j_0$, we get the two-dimensional distribution $p(j_1,j_0)$. This distribution
actually satisfies fluctuation theorem in Equation~(\ref{eq5}), as illustrated in
Figure~\ref{fig5}c. Let us emphasize that the transition to the variables $j_1$ and $j_0$ is not the transition to the principal axes of the distribution $p(j_1,j_2)$.

\begin{figure}[h!]
	\centering	
	\includegraphics[scale=0.7]{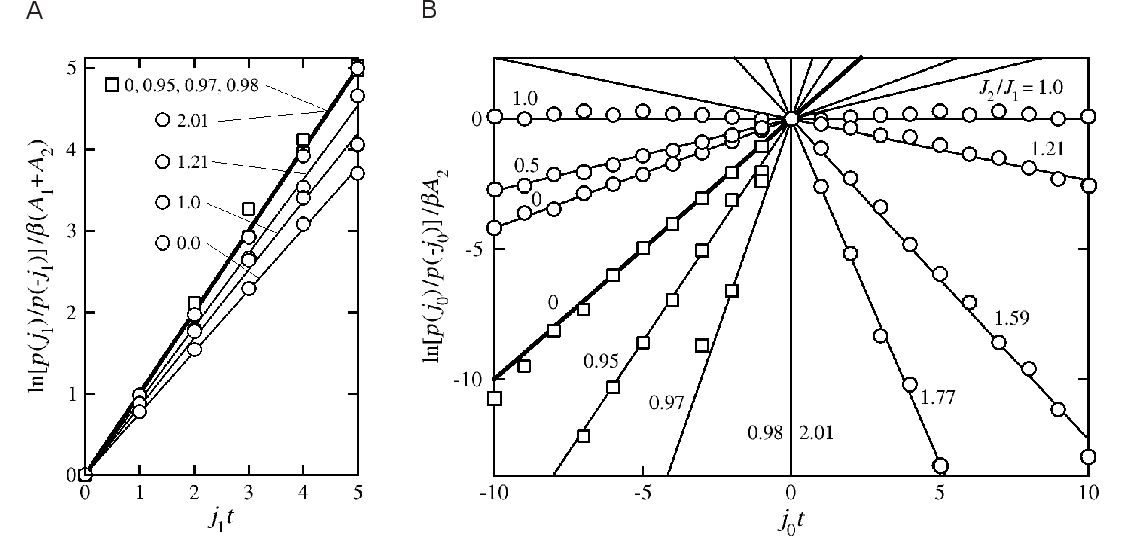}
	\caption{The generalized fluctuation theorem dependences found in the random walk simulations
		on the network shown in Figure~\ref{fig3}e with the single output gate (the squares) and the
		fourfold output gate (the circles). We assumed $\beta A_1 = 1$ and a few smaller, negative
		values of $\beta A_2$ determining the ratio of averaged fluxes $\epsilon = J_2/J_1$ noted in
		the graphs. (a)~The case of marginal $p(j_1)$, compare Equation~(\ref{eq10}). (b)~The case of
		marginal $p(j_0)$, compare Equation~(\ref{eq11}). In order to clearly distinguish the
		contributions from energy dissipation and information, the results were divided by
		dimensionless forces $\beta (A_1 + A_2)$ and $\beta A_2$ for $i =$ 1 and 0, respectively. In
		this way, the bold lines in both graphs, with unit tangent of the inclination angle,
		correspond to the dissipation components. Let us recall that $\beta (A_1+A_2)$ is positive
		while $\beta A_2$ is negative.} 
	\label{fig6}
\end{figure}

From two-dimensional distributions $p(j_1,j_0)$, we calculated the marginal distributions $p(j_1)$
and $p(j_0)$ for all simulated trajectories. The logarithms of the ratio of marginals
$p(j_i)/p(-j_i)$ are presented in Figures~\ref{fig6}a and b as the functions of $j_{i}t$ for $i =
1$ and $0$, respectively. All the dependences are actually linear as predicted by generalized
fluctuation theorems in Equations~(\ref{eq10}) and (\ref{eq11}). 

For the single output gate (the squares in Figure~\ref{fig6}), we got both $\iota_1$ and $\iota_0$
negative, adding up to positive $\sigma_1$ and $\sigma_0$. This is consistent with the determined
negative values of correlation coefficient $\rho$. Both information contributions differ from zero
only for $\epsilon$ close to unity, i.e. for very small values of force $A_2$, which does not
break practically the detailed balance condition for the transitions through the output gate and
are almost a usual white noise. In consequence, information $\iota_1$ is unnoticeable in
Figure~\ref{fig6}a ($A_1 + A_2 \approx A_1$), and the information $\iota_0$ in Figure~\ref{fig6}b,
determined mainly by the flux through the single output gate, becomes visible only after dividing
by a small value of $\beta A_2$. 

For the fourfold output gate (the circles in Figure~\ref{fig6}), the situation is much more
interesting. Now, information contributions $\iota_i$ are of the same sign as dissipation
contributions $\sigma_i$ and substract from them, which is consistent with the positive values of
correlation coefficient $\rho$. Dissipations $\beta D_i = \langle\sigma_i\rangle$ and informations $I_i = \langle\iota_i\rangle$ for $i =1$ and $0$,
corresponding to the inclination of the straight lines in Figure~\ref{fig6}b, are presented in
Figure~\ref{fig7} as the functions of output force $\beta A_2$ determining the value of the degree
of coupling $\epsilon = J_2/J_1$. For $J_2/J_1 < 1$, which corresponds to large values of output force $\beta A_2$,
information $I_0$ is positive, i.e. taken from the microdynamics, to be next subtracted from the
positive dissipation $\beta D_0$. For $J_2/J_1 > 1$, which corresponds to small values of output
force $\beta A_2$, information $I_0$ is negative, i.e. sent to the microdynamics, and dissipation
$\beta D_0$ becomes negative. So we showed that the organizational subsystem in our model actually
behaves like Maxwell's demon.

\begin{figure}[h!]
	\centering
	\includegraphics[scale=0.5]{fig7}
	\caption{Contribution of dissipation $\beta D_i$ and information $I_i$ to observed entropy
	production for the studied model system with the fourfold output gate. $i = 1$ in the case of
	operation energy processing and $i = 0$ in the case of organization energy processing, compare
	Figure~\ref{fig4}b. All quantities are presented as function of the ratio of forces $A_2/A_1$
	or fluxes $J_2/J_1$ and counted in nats per random walk step.}
	\label{fig7}
\end{figure}

\section{Conclusions and suggestions for further research}

Under physiological conditions, the protein molecular machines fluctuate constantly between the
multitude of conformational substates that make up their native state. The probabilities of
visiting individual substates are far from the equilibrium and determined by the concentration of
the surrounding molecules involved in the process. The protein machines are bifunctional enzymes
that simultaneously catalyze two processes which can be considered as two effectively unimolecular
chemical reactions: the input reaction donating free energy and the output reaction accepting it.
During the full cycle of free energy-transduction, the possibility to choose different ways of the
free energy-accepting reaction results in the transient limitation of the dynamics to different
regions of the conformational network, that is, to breaking the ergodicity \cite{Engl06,Metz14}. The transient ergodicity breaking makes the machine's internal dynamics to be a memory for storing
and manipulating information. Information is erased each time the energy-donating reaction starts
the next cycle. The storage capacity of memory is higher the larger and more complex is the
network of conformational substates. This network is particularly large in the case of protein
motors, since it then contains the substates of the motor and the whole track on which it moves
\cite{Kurz06}, compare Figures~\ref{fig3}c and \ref{fig4}e.

Information is transferred between constrains and memory of the machine in the form of Shannon's entropy, an addition to free energy. The two free energy components determine the thermodynamic state of the machine. The
first, proportional to the concentration of the input reaction molecules $X_1$, we refer to as the
operation energy. And the second, proportional to the difference in the concentrations of the
output and input reaction molecules $X_2-X_1=X_0$, we refer to as the organization energy.
According to the generalized fluctuation theorems that apply for each variable separately,
information, like work and dissipation, is a change in these well defined functions of state.
Dissipation $D_1$ and information $I_1$ are associated  with the variable $X_1$, and dissipation
$D_0$ and information $I_0$ are associated with the variable $X_2-X_1$. If the nanomachine has no
choice possibility, both $D_1$ and $I_1$, as well as $D_0$ and $I_0$ are of opposite signs, and
information processing results in an additional free energy loss. Only in the nanomachines that
choose randomly the way of doing work and transmit information about it to microdynamics, $D_0$
and $I_0$ can be negative, which makes their organizational subsystem to be Maxwell's demon. Of
course, according to the fluctuation theorem, Equation~(\ref{eq6}), the resultant entropy
production in both subsystems must always be non-negative. 

Under macroscopic stationary conditions, the constant values of the both free energy components
can be assigned to each individual protein machine. Fluctuating physical quantities are fluxes
exerted by corresponding forces, specified for a certain period of time $t$, not too long and not
too short. The probability distributions of these fluxes determine average dissipations $\beta
D_i$ in $k_{\rm B}T$ units and informations $I_i$. In Figure~\ref{fig7}, which presents the most
important result of the paper, two values of the ratio of average fluxes $\epsilon = J_2/J_1$
correspond to the absence of dissipation $D_0$. 

The first is for the case of tight coupling, $\epsilon = 1$, i.e. $J_0 = J_2 - J_1 = 0$, for which
the zero free energy dissipation $D_0$ occurs simultaneously with the zero information transferred
$I_0$. The necessary condition for it is a special, critical value of output force $A_2$
distinguished in Figure~\ref{fig7} by the vertical dashed line. There are some arguments that,
effectively, such the force value is realized for processive motors with a feedback control
between their two identical components \cite{Caop04,Bier07,Yild08}. The information transfer to
memory takes place for the force of a value less than critical. Only one of the monomers is
affected by this force, but the other one, indifferent to external interactions, observes the
first monomer and, depending on the result of the observation, exerts an appropriate additional
force on it, so that the resulting force reaches the critical value. In consequence, the chosen
monomer works as a tightly coupled perfect machine. The process is repeated with the conversion of
one monomer into another. The suggested mechanism seems to be in harmony with the consensus
kinesin-1 chemo-mechanical cycle \cite{Hanc16} and, in a broader perspective, a closer examination
of it could help answer the certainly interesting question: why do most protein machines operate
as dimers? 

The second case of zero dissipation $D_0$ takes place for $\beta A_2$ tending to 0, when the
system is no longer working as a machine, but dissipation $D_1$ is used for collecting information
$I_0$. Such conditions are met by the transcription factors in their one-dimensional search for
the target on DNA, intermittent by three-dimensional flights \cite{Kolo11}. According to our
theory, the flights should correspond to the randomly chosen external transitions, hence the
one-dimensional sliding between them may not be simple diffusion but the active continuous time
random walk \cite{Metz14}. In fact, the external transitions, assumed in the present model to be
very fast, are also continuous time random walk \cite{Lium17}. During information collection
ergodicity is broken, which agrees with the assumption that the macromolecule of the transcription
factor is derived from the equilibrium throughout the entire search period.

Let us finish with a general remark. The generalized second laws of thermodynamics in
Equations~(\ref {eq17}) were derived only for isothermal processes in nanoscopic machines. Their
universality remains an open problem. One thing is for sure. To get the Equations~(\ref{eq17}),
three conditions were necessary: (i) openess of the system providing an external source
of free energy and steady-state dynamics, (ii) characterization of the system by an organizational
variable, and (iii) fluctuation of fluxes. 

\section*{Author Contributions}
The general concept and the theory is mainly due to M.\,K., who also wrote the manuscript. The
specification of the critical branching tree model and the numerical simulations are mainly due
to P.\,C.

\section*{Conflicts of Interest}
The authors declare no conflict of interest.

\section*{Acknowledgments}
M.\,K. thanks Ya\c{s}ar Demirel and Herv\'{e} Cailleau for discussing the problem in the early
stages of the investigation.

\bibliographystyle{unsrt}  
\bibliography{infoeuro}  

\end{document}